\pdfoutput=1

\documentclass[aps,twocolumn,showpacs,preprintnumbers,amsmath,amssymb]{revtex4}

\usepackage{graphicx}

\begin{document}

\title{Contractile prestress controls stiffening and fluidization of living cells\\ in response to large external forces}

\author{Philip Kollmannsberger}
\author{Claudia T. Mierke}
\author{Ben Fabry}

\affiliation{Department of Physics, University of Erlangen, D-91052 Erlangen, Germany}

\date{\today}

\begin{abstract}
We report the simultaneous characterization of time- and force-dependent mechanical properties of adherent cells in the physiologically relevant regime of large forces. We used magnetic tweezers to apply forces to magnetic beads bound to the cytoskeleton, and recorded the resulting deformation (creep response). The creep response followed a weak power law at all force levels. Stress stiffening and fluidization occurred simultaneously and were quantified by the force-dependence of the creep compliance and the power law exponent. The amount of stiffening and fluidization in response to force was controlled solely by the contractile prestress of the cell and followed a simple relationship. This behavior may be of fundamental importance in biological processes that involve a mechanical interaction between cells and their environment.\end{abstract}


\maketitle

Common human disorders such as cancer, inflammatory or cardiovascular diseases are often associated with derangements of cell rheological properties \cite{Suresh2005}. Experimental advances in microrheology during the past years revealed that the rheology of living cells can be summarized by a few simple empirical relationships \cite{Bausch1998,Yamada2000,Fabry2001,Alcaraz2003,Trepat2008}. Accordingly, the linear creep response $J(t)$ and  dynamic shear modulus $G(\omega)$ under small deformations follow a weak power law over several orders of magnitude in time or frequency. Moreover, the stiffness increases linearly with contractile cell tension (cytoskeletal prestress), as observed under pharmacological modulation of myosin motor activity and simultaneous measurement of the linear shear modulus \cite{Wang2001}. The increase of stiffness with cytoskeletal prestress as well as the power-law viscoelastic moduli have been characterized using linear microrheology. However, the large forces and deformations that cells experience under physiological conditions in the living organism often exceed the linear regime \cite{Fung1993}. Cells have been reported to stiffen, to soften, or to fluidize after application of external stretch, depending on experimental conditions and cell models used \cite{Fernandez2008,Bursac2005,Trepat2007}. A comprehensive description of the nonlinear mechanical behavior of cells is currently missing. In particular, simultaneous measurements of time- and force-dependent nonlinear properties in response to large forces are needed to understand how stress stiffening, softening and fluidization contribute to the nonlinear rheology of cells.

In this letter, we report measurements of the microrheological creep response of living cells to large forces. A staircase-like sequence of increasing force steps was applied to magnetic beads bound to the cytoskeleton. The resulting bead displacements were recorded and analyzed using a nonlinear superposition approach. We found power-law time dependence of the creep response regardless of the applied force, and an increase of both the stiffness and the power law exponent with force. The amount of stress stiffening was smaller for stiffer cells. We attribute this to a smaller relative increase of the cytoskeletal (internal) stress after external force application in cells with higher prestress. Furthermore, the increase of the power law exponent was smaller for more fluid-like cells. This suggests that the turnover rate of cytoskeletal bonds is less sensitive to external force in these cells. Our results show that both in the linear and nonlinear regime, elastic and dissipative properties are controlled by a balance of internal and external stress, and that living cells can actively tune their stress stiffening and fluidization behavior in response to large external forces through their contractile prestress.

\begin{figure}
\includegraphics{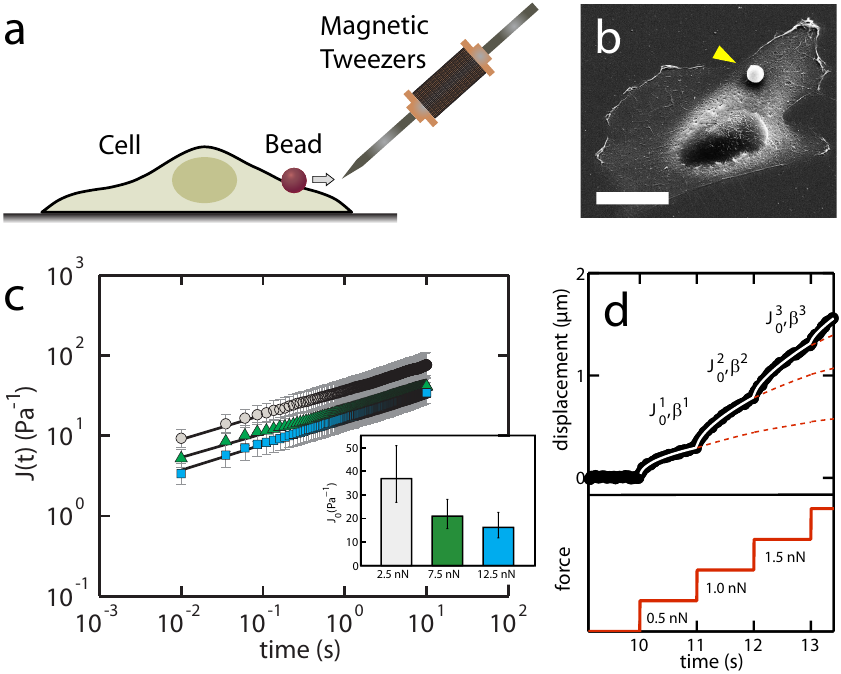}
\caption{\label{Fig1} (a) The force generated by Magnetic Tweezers acts on superparamagnetic beads (Dynabead M-450, Invitrogen) coated with fibronectin (FN, Roche Diagnostics) which are bound to the cytoskeleton of adherent cells via integrin receptors. (b) SEM image of a fibroblast with a 4.5 $\mu$m bead (arrow) bound to its surface. Bar = 20 $\mu$m. (c) Creep response to force steps of 2.5 nN (top), 7.5 nN (middle) and 12.5 nN (bottom) always followed a power law over time, $J(t)=J_0(t/t_0)^\beta$. Inset: The decrease of the prefactor $J_0$ with increasing force indicates stress stiffening. (d) A staircase-like sequence of increasing force steps was applied, and the displacement was fit to a superposition of creep processes.}
\end{figure}

To measure the creep response, we used a magnetic tweezers setup as described in \cite{Kollmannsberger2007} that was optimized for applying high forces in the 10 nN-range to magnetic beads bound to living cells. Superparamagnetic 4.5 $\mu$m beads (Dynal, Invitrogen) were coated with fibronectin (FN) (100 $\mu$g/ml, Roche Diagnostics). Prior to measurements, the FN-coated beads were sonicated, added to the cells and incubated for 30 min. Seven different cell lines were measured (mouse embryonal fibroblasts, NIH 3T3 mouse fibroblasts, F9 mouse embryonic carcinoma cells, MeWo human fibroblast-like cells, and MDA-MB231, 786-O and A125 human epithelial cancer cells). A single force step or a staircase-like increasing force was applied for 10 seconds. The resulting bead displacement $d(t)$ was determined from images recorded with a CCD-camera (Orca-ER, Hamamatsu) at a rate of 40 frames/s. Bead displacements during 10 seconds prior to force application were also measured to correct for stage drift, cell migration and directed bead motion during measurements.

Bead displacement $d(t)$ during a constant force pulse $f_0$ followed a weak power law, $d(t)\propto t^\beta$. This power-law time dependence was independent of the applied force magnitude (Fig. 1). We estimate the typical strain $\gamma(t)$ as the displacement $d(t)$ divided by the bead radius $r$, and the typical stress $\sigma$ as the applied force divided by the bead cross-sectional area, $r^2\pi$. The creep compliance $J(t)$ in units Pa$^{-1}$ is then given by $d(t)/f$ times a constant geometric factor $\pi r\approx 7.1$ $\mu$m and is fitted to the equation
\begin{equation}
\label{powerlaw}J(t)=J_0\left(t/t_0\right)^\beta
\end{equation}
with time normalized to $t_0=1$ s. The power-law exponent $\beta$ characterizes the time-dependent viscoelastic properties and was between 0.1 and 0.5, where $\beta=0$ corresponds to an elastic solid and $\beta=1$ to a viscous fluid. The inverse of the prefactor, $1/J_0(\sigma)$, is equivalent to a differential shear modulus $K'(\sigma)$ at time $t=1$ s. The creep response to force steps of different magnitude always followed a power law. $J_0$ decreased with force (Fig. 1c), indicating stress stiffening and the breakdown of linear superposition.

In order to quantify the force dependence of the creep response, a staircase-like sequence of increasing force steps was applied. The cell strain after the n-th force step $\sigma_n$ at time $t_n$ was fitted to a superposition of creep processes,
\begin{equation}
\label{step}
\gamma(t\geq t_n)=\gamma(t_n)+\sum_{i=0}^n \left[J_\sigma(t-t_i)-J_\sigma(t_n-t_i)\right]\sigma_i,
\end{equation}
with a response function $J_\sigma(t)$ that depends not only on time but also on the currently applied total force:
\begin{equation}
\label{forcepowerlaw}
J_\sigma(t)=J_0(\sigma)\left(t/t_0\right)^{\beta(\sigma)}.
\end{equation}

Stress stiffening (increase of $1/J_0$ with force) as well as fluidization (increase of $\beta$ with force) were observed in all cell types. Fibroblasts were on average stiffer but showed less stress stiffening compared to epithelial cells. To quantify the relationship between stiffness and stress stiffening, data from all experiments were pooled and binned by stiffness. The stiffest and most elastic cells showed the smallest amount of stress stiffening (Fig. 2).

In the following, we argue that different degrees of stiffening are caused by different levels of prestress in the cell. We assume that the total mechanical stress $\sigma$ in the cytoskeleton is the sum of active (myosin-generated) internal prestress $\sigma_p$ and passive external stress $\sigma_e$. Furthermore, we assume that the linear relationship between the differential stiffness $K'(\sigma)$ and cytoskeletal stress, which has been previously reported for smooth muscle cells \cite{Wang2001} and reconstituted actin networks \cite{Gardel2006}, is a universal property and also holds in the cell lines tested here. $K'(\sigma)$ can then be expressed as
\begin{equation}
\label{stiffnessprestress}
K'(\sigma)=\frac{d\sigma}{d\gamma}=K'_0+a(\sigma_p+\sigma_e),
\end{equation}
where the unitless constant $a$ characterizes the dependence of stiffness on stress, and $K'_0$ denotes the linear stiffness around the force-free state. Integration yields an exponential stress-strain relationship, as known for whole cells, reconstituted cytoskeletal networks and many other biological tissues \cite{Fung1993,Fernandez2008,Storm2005}.

We fitted the force-stiffness curves in Fig. 2a to eq. \eqref{stiffnessprestress} and found that the different degrees of stiffening are explained solely by $\sigma_p$. Accordingly, stiff cells have a more prestressed cytoskeleton, therefore the relative increase of mechanical tension and resulting stress stiffening due to external forcing is smaller than for soft cells. The values of $\sigma_p$ obtained from the fit are proportional to the average measured stiffness $K'_0$ at the smallest force $\sigma_0$ (Fig. 2, inset). The fit yields prestress values of up to 1500 Pa, which is in agreement with the maximum traction stress these cells exert on their substrate \cite{Wang2001}. $K'_0$ corresponds to the stiffness of the unstressed and prestress-free cell. Interestingly, the value of 5 Pa for $K'_0$ obtained from the fit is similar to the linear shear modulus of crosslinked actin networks \cite{Gardel2006}.

In order to test whether different degrees of stress stiffening are due to differences in the actomyosin-generated prestress, we inhibited contraction with the myosin light chain kinase (MLCK) inhibitor ML-7 in the highly contractile mouse embryonal fibroblast (MEF) cells. As predicted by eq. \eqref{stiffnessprestress}, a reduction of prestress by ML-7 results in reduced stiffness and a more pronounced stress stiffening (Fig. 2, inset).

\begin{figure}
\includegraphics{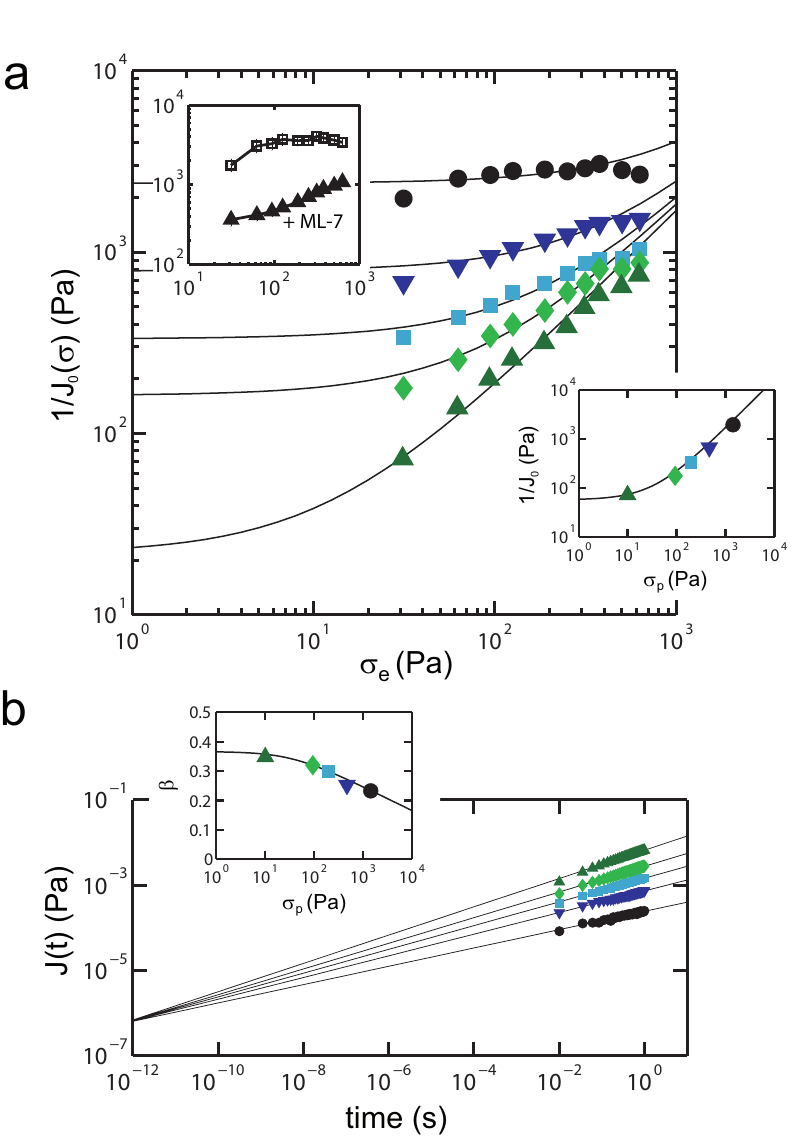}
\caption{\label{Fig2} (a) Stiffness $1/J_0(\sigma)$ vs. applied stress $\sigma_e$, binned by the stiffness at the smallest force (n = 395). Stiffer cells show less stress stiffening. Black lines: fit to eq. \eqref{stiffnessprestress} with parameters $a = 1.68$ and $K'_0 = 5$ Pa, and prestress $\sigma_p$ as free parameter. Top inset: Inhibition of cytoskeletal prestress by ML-7 in MEF cells (black triangles) reduced stiffness and accentuated stress stiffening compared to untreated cells (open squares). Lower inset: Measured initial stiffness $1/J_0$ vs. $\sigma_p$ for all data. Solid line: prediction by eq. \eqref{stiffnessprestress}. (b) Creep response $J(t)$ for the first force step vs. time, same binning as in (a). Solid lines: fit to eq. \eqref{master}, with parameters $j_0=5.59\times 10^{-7}$ Pa$^{-1}$ and $\tau_0=5.5\times 10^{-13}$ s, and $\beta$ as free parameter. Inset: measured exponent $\beta$ vs. $\sigma_p$ from Fig. 2a. Solid line: prediction by eq. \eqref{b-prestress}. Standard errors are smaller than symbol size in all cases.}
\end{figure}

The stiffness-binned creep curves during the first (0.5 nN) force step exhibit a common intersection at small times (Fig. 2b) and can be fit to the scaling equation \cite{Fabry2001}

\begin{equation}
\label{master}
J(t)=j_0\left(t/\tau_0\right)^\beta
\end{equation}

with a common intersection at $j_0=5.59\times 10^{-7}$ Pa$^{-1}$ and $\tau_0=5.5\times 10^{-13}$ s. As has been suggested previously \cite{Stamenovic2004}, for eq. \eqref{stiffnessprestress} and eq. \eqref{master} to hold at the same time, the following relationship between prestress, stiffness and power law exponent must also hold:

\begin{equation}
\label{b-prestress}
\beta(\sigma_p)=\frac{\ln\left[j_0\left(K'_0+a\sigma_p\right)\right]}{\ln\tau_0}.
\end{equation}

This means that more contractile cells (higher prestress) are not only stiffer, they also display a smaller power law exponent and hence more solid-like properties compared to less contractile cells. The creep exponent obtained from the data in Fig. 2b, when plotted against the prestress obtained from the data in Fig. 2a, closely follows eq. \eqref{b-prestress} (Fig. 2b, inset).

\begin{figure}
\includegraphics{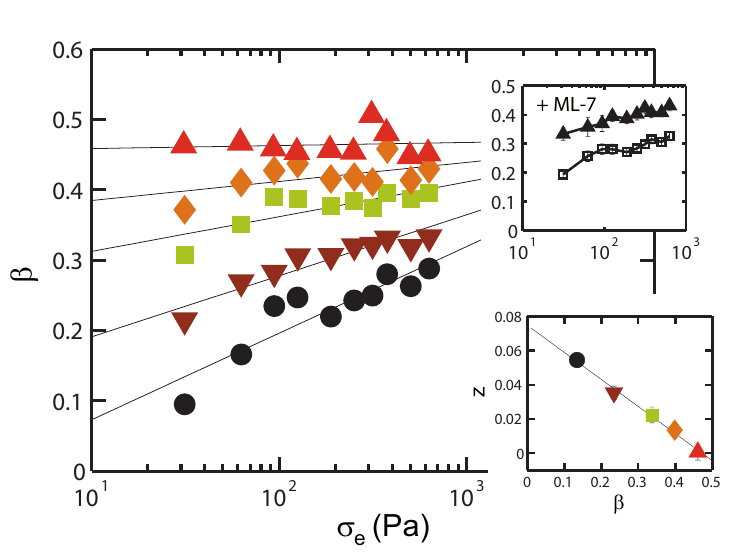}
\caption{\label{Fig3} Power law exponent $\beta$ vs. applied stress $\sigma_e$. Data for all cell lines are binned by the exponent at the smallest force, $\beta_0$ (n = 395). Elastic cells (small $\beta_0$) show a higher amount of fluidization in response to force application compared to fluid cells (large $\beta_0$). Solid lines: fit to the empirical relation $\beta(\sigma_e) = \beta_0 + z\log(\sigma_e/\sigma_0)$. Top inset: Inhibition of cytoskeletal prestress by ML-7 in MEF cells (black triangles) leads to an increase in $\beta$ but reduces fluidization compared to untreated cells (open squares). Lower inset: The amount of fluidization, $z$, decreases linearly with the initial exponent $\beta_0$ (solid line).}
\end{figure}

This relationship (eq. \eqref{b-prestress}), however, does not predict the behavior of $\beta$ for higher external forces. The power law exponent $\beta$ increased during force application, indicating force-induced fluidization and yielding events (Fig. 3). Cells with more solid-like behavior (small $\beta$) showed the most pronounced fluidization, whereas cells that were initially more fluid-like (large $\beta$) showed no further increase of $\beta$ during creep.  Inhibition of MLCK by ML-7 in MEF cells reduced cellular prestress, resulting in an increase of $\beta$ and reduced fluidization (Fig. 3, inset).

Taken together, these data show that the linear (stiffness and power law exponent) and the nonlinear (stress stiffening and fluidization) mechanical properties of cells are controlled by the cytoskeletal prestress. We ruled out that stiffening and fluidization were caused by an activation of mechanosensory pathways in response to external force application.  By repeating the creep experiments with poly-L-lysine-coated beads that bind to the cytoskeleton through non-specific transmembrane proteins, we avoided integrin receptor activation but still observed stress stiffening (data not shown). Other active stiffening mechanisms such as contraction due to stretch-induced ion channel activation can be ruled out by the fact that the stiffening response is instantaneous (Fig. 1c), time-independent (lower insets Fig. 2a, 3a), and is also seen after inhibition of myosin contraction with ML-7 (Fig. 2, 3). To exclude stiffening due to active remodeling of the cytoskeleton on longer timescales, we completed our experiments within ten seconds.

Power law rheology, passive stress stiffening and direct proportionality between stiffness and internal or external prestress are ubiquitous in biopolymer networks \cite{Koenderink2009,Storm2005,Gardel2006}. These observations together with our data support the long-held notion that the same physical principles that govern the rheology of semiflexible filament networks also apply to the cytoskeleton of living cells, and that stress stiffening and power-law rheology have a structural rather than a molecular origin.

Power-law rheology is the macroscopic footprint of a broad distribution of relaxation time constants of the underlying microscopic dissipation mechanism, as described by the theory of Soft Glassy Rheology (SGR) \cite{Sollich1997}. SGR has been a useful concept for understanding cell mechanical behavior \cite{Fabry2001} and has generated a large number of predictions that have been experimentally confirmed \cite{Bursac2005,Trepat2007}. While SGR is in good agreement with the power-law behavior and fluidization that we find, it does not account for stress stiffening. This limitation has recently been overcome by combining the SGR concept with models of semiflexible polymer networks \cite{Semmrich2007}.

Neither SGR nor semiflexible polymer rheology can explain the observation that the power-law exponent decreases in cells with higher prestress but increases in response to higher external stress (Fig. 3). Similar behavior has previously been observed in smooth muscle and has been explained by the ,,latch hypothesis`` where the more solid-like behavior under high prestress arises from reduced actomyosin cycling and energy dissipation due to a strongly bound state of actin and myosin \cite{Fredberg1996}. An alternative explanation is that cytoskeletal bonds deviate from the typical Bell-type ,,slip-bond`` behavior and instead show increased lifetimes under force \cite{Guo2006}. 

Regardless of mechanism, our results demonstrate that adherent cells control the amount of stress stiffening and fluidization in response to large external forces by modulating their contractile prestress. More contractile cells are not only stiffer and more solid-like, they also show less stress stiffening and increased fluidization compared to less contractile cells. The biological relevance of this behavior is that soft and liquid-like cells, on the one hand, need to stiffen in order to withstand large mechanical stress. Stiff and solid-like cells, on the other hand, depend on fluidization in order to be able to withstand large mechanical strain without rupturing. Thus, by a single mechanism that is present in all eukaryotic cell types –- namely, by modulating the activity of actomyosin contraction -– cells can adapt to a wide range of mechanical conditions. Our data show that the time-dependent mechanical responses of adherent cells to large forces and deformations obeys clearly defined empirical laws. Their knowledge is important for a quantitative understanding of biological processes involving mechanical interaction between cells and their environment, such as matrix remodeling, mechanosensing, cell migration or tissue development.

We thank K. Kroy, P. Fernandez and J.P. Butler for helpful discussions. This work was supported by NIH, DFG and Deutsche Krebs\-hilfe.

\end{document}